\documentclass[american,superscriptaddress,aps,reprint,pre]{revtex4-2}

\usepackage[T1]{fontenc}
\usepackage[utf8]{inputenc}
\usepackage{babel}
\usepackage{amsmath}
\usepackage{amssymb}
\usepackage{bm}
\usepackage{graphicx}
\usepackage{microtype}
\usepackage{etoolbox}
\usepackage{xcolor}
\usepackage{mathtools}
\usepackage[colorlinks=true,citecolor=blue,linkcolor=blue,urlcolor=blue]{hyperref}
\newcommand{\defeq}{\vcentcolon=}

\newcommand{\Gini}{\mathcal{G}}
\newcommand{\lamm}{{\lambda_\mathrm{m}}}
\newcommand{\ym}{{y_\mathrm{m}}}
\newcommand{\ymax}{{y_\mathrm{max}}}
\newcommand{\Phim}{{\Phi_\mathrm{m}}}
\newcommand{\tsc}{{t_\mathrm{sc}}}
\newcommand{\Nsc}{{N_\mathrm{sc}}}
\newcommand{\xsc}{{x_\mathrm{sc}}}
\usepackage{enumitem}

\usepackage{tabularx}

\apptocmd{\thebibliography}{\sloppy}{}{}
\begin{document}

\title{Self-similar Inequality Dynamics in Biased Competition for a Conserved Resource}

\author{Giacomo Bartolucci}
\affiliation{Department of Condensed Matter Physics, Universitat de Barcelona, 08007 Barcelona, Spain}
\affiliation{Universitat de Barcelona Institute of Complex Systems (UBICS), Universitat de Barcelona, 08028 Barcelona, Spain}

\author{Jonathan Bauermann}
\affiliation{Department of Physics, Harvard University, Cambridge, Massachusetts 02138, USA}

\date{\today}

\begin{abstract}
In many complex systems the redistribution of a conserved resource can generate strong inequality. Important examples are cities exchanging inhabitants, farms competing for farmland, or agents for wealth in saturated markets.
Here we model such systems as agents competing for a fixed total resource, and show that coupling agent extinction to a bias favoring richer agents dynamically amplifies this inequality over time. Drawing on an analogy with emulsions, where droplets compete for a conserved quantity, we introduce a dynamical critical resource size that determines an agent's fate. 
Crucially, this threshold grows in time, continuously raising the bar for survival and further reinforcing inequality. 
Increasing the bias strength, we find a transition from a continuous resource concentration towards a winner-takes-all outcome.
Furthermore, the case of continuous resource concentration is characterized by a self-similar dynamics accompanied by a power-law or exponential increase in inequality, quantified by the Gini coefficient. 
We derive these results analytically in a mean-field model and compare them to agent-based simulations, showing excellent agreement.

\end{abstract}

\maketitle

\section{Introduction}

Ever since Pareto's observation that wealth concentrates in the hands of a few, understanding how inequality emerges has been a defining challenge across the social sciences~\cite{paretoCoursDeconomiePolitique1896,yakovenkoColloquiumStatisticalMechanics2009}.
During the period known as the \textit{Great Acceleration}~\cite{Steffen2015}, the explosive rise in production and consumption motivated the study of how heterogeneous growth translates into inequality.
A first attempt is Gibrat's law, which states that the absolute growth rate of market participants (agents) is linearly proportional to their sizes~\cite{gibratInegalitesEconomiquesApplications1931, suttonGibratsLegacy1997}.

A pivotal framework for the origin of inequality in expanding systems was proposed by Bouchaud and Mézard~\cite{Bouchaud2000}, who modeled heterogeneous growth through a multiplicative noise term acting on each agent's resource. This produces a steady-state distribution with power-law tails, the celebrated Pareto distribution, which has been reported in a wide range of contexts such as city populations and firm sizes~\cite{gabaixPowerLawsEconomics2009, gabaixZipfsLawCities1999, axtellZipfDistributionUS2001}.

Though many economies are described in terms of expansion, not every relevant system can grow indefinitely: many are constrained by land, attention, market share, or other slowly varying factors.
In such saturated systems or fixed pools, agents compete over an approximately conserved resource, so that the growth of one is necessarily coupled to the decline of another.

The absence of net growth does not preclude inequality: on the contrary, even ``fair'' trades can generate it, provided the total resource being exchanged is conserved. This phenomenon is captured by \textit{metaphorical models}~\cite{Bouchaud2023} of random, unbiased asset exchange such as the yard-sale model~\cite{CHAKRABORTI2002}. 
These models show that conservative, unbiased exchange alone can lead to wealth condensation generating strong inequality~\cite{Boghosian2014},
because poorer agents risk a larger fraction of their wealth and random fluctuations tend to concentrate wealth among the rich. Yard-sale models have also been generalized to include wealth redistribution and a bias favoring the wealthier agents. The competition between these two factors leads to a stationary wealth distribution that can undergo a phase transition to a state in which a finite fraction of the wealth condenses onto a single agent~\cite{Boghosian2017}. 

Hitherto, most work on both growing and stagnating economies has focused on the tails of steady-state distributions. Far fewer studies have addressed systems that never reach a steady state, where inequality keeps rising~\cite{ispolatov1998,leyvraz2002,Ben-Naim2003,complexis16}.
Elaborating on these seminal works, we study biased exchanges of a conserved resource and take a different route: we characterize the evolving resource distribution through a critical size, the time-dependent scale that separates growing agents from shrinking ones. We derive how this scale evolves and connect its dynamics, together with that of the Gini coefficient, to the dynamics of inequality.

Explicitly, we focus on saturated systems with a bias favoring the agents holding more resource such that the rich get richer, and, because exchanges are conservative, the poor get poorer, amplifying small differences over time. We further allow bankruptcy, so that agents can drop out of the competition, and their number is itself not conserved, adding a second, coupled layer of dynamics to the resource redistribution.

\begin{figure*}[t]
    \centering
    \includegraphics[width=0.99\linewidth]{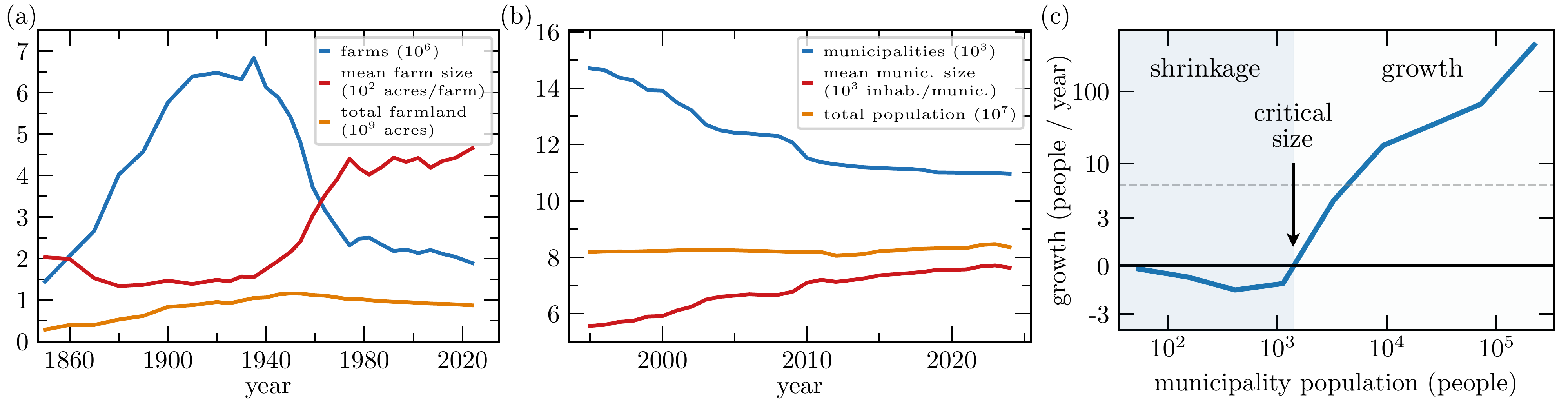}
    \caption{(a) Number of US farms, mean farm size, and total US farmland~\cite{USDAFarms} in the years between 1850 and 2024. (b) Number of German municipalities, average municipality population, and total population of Germany between 1995 and 2024~\cite{RegionaldatenbankDeutschland2026}. (c) Growth of German municipalities, binned by size with at least 20 municipalities per bin, averaged over the years 2000-2022. For better visualization, the y-axis above the gray dashed line is log-scaled.
   }
    \label{fig:data}
\end{figure*}

Real-world examples of this combination, i.e., a bias in the competition, a conserved resource, and agent extinction, abound. An emblematic case is agricultural land in the United States: during the long westward expansion, both the land mass used for farming and the number of farms constantly increased, see Fig.~\ref{fig:data} (a). Since the middle of the 20th century, however, the land area has remained approximately constant, while bankruptcy and mergers have driven the number of farms down such that the average farm size increased~\cite{USDAFarms}, a trend also visible in the number of sold cars and car dealerships in the US since the late 1980s~\cite{FREDTOTALSA,DOEFact695}.

In Germany, the same reasoning applies to cities exchanging inhabitants: the total population has been approximately constant over the last three decades, while shrinkage and merging have steadily reduced the number of municipalities, so that, by population conservation, the average number of inhabitants per municipality has increased; see Fig.~\ref{fig:data}(b). Inspecting the size-dependent growth in Fig.~\ref{fig:data}(c) reveals a critical municipality size at which the growth curve crosses zero. This indicates the existence of a critical size above which municipalities grow, while the ones below shrink.

Coarsening by ripening offers a natural physical analogy for this behavior. Indeed, the presence of a critical size and the decay of the agent number concomitant with the rise of average agent resource is reminiscent of Ostwald ripening. During Ostwald ripening, droplets or clusters exchange a conserved quantity and surface tensions drive larger clusters to grow at the expense of smaller ones~\cite{voorhees1985theory}. The result is a dynamical scaling in which the number of clusters decreases, the typical cluster size grows, and the distribution approaches a self-similar form~\cite{Carr1998,Niethammer1999}, as described by the Lifshitz-Slyozov-Wagner (LSW) theory~\cite{Lifshitz1961,Wagner1961}. The ripening picture has already inspired models in the social sciences, from urban segregation~\cite{schellingDynamicModelsSegregation1971,vinkovicPhysicalAnalogueSchelling2006} to opinion dynamics~\cite{dallastaAlgebraicCoarseningVoter2008, caccioliVoterModelsConserved2013, MUNJAL2019182}. Furthermore, in a seminal work, Leyvraz and Redner showed that migration-driven aggregate growth, in which
members move preferentially from smaller to larger aggregates, admits a scaling
theory closely corresponding to LSW coarsening, and proposed applications to the growth of city populations and to wealth exchange~\cite{leyvraz2002}.

Here, motivated by these observations, we propose a minimal agent-based model of biased competition for a conserved resource in which the number of agents is not conserved, and introduce a mean-field description in terms of a critical resource amount. We then analytically derive the scaling in time and the collapse of the resource distribution at late times, in the spirit of LSW theory.
Unlike classical ripening in an emulsion, where the competitive advantage
weakens as coarsening proceeds, we study an advantage scaling sublinearly,
linearly, or superlinearly with size.
This determines both the scaling of the critical size and the qualitative behavior of the system, ranging from continuous concentration of wealth to a finite-time winner-takes-all dynamics.

\section{Biased Asset-Exchange Model}
\label{sec:agent_model}

We first introduce a conservative agent-based model. Initially, there are $N_0$ agents each holding a nonnegative resource $x_i(t)\geq 0$, with fixed total resource
\begin{equation}
    R=\sum_{i=1}^{N_0} x_i(t).
\end{equation}
At each microscopic step, active agents are randomly paired. For a pair $(i,j)$, we define the uncapped transfer
\begin{equation}
    \Delta_{ij}=\eta\left(x_i^\alpha-x_j^\alpha\right),
    \label{eq:microscopic_transfer}
\end{equation}
where $\eta>0$ is an exchange scale and $\alpha$ controls the strength of the bias. The actual transfer is capped by the resource held by the losing agent,
\begin{equation}
    T_{ij}=
    \begin{cases}
    \min\{\Delta_{ij},x_j\}, & \Delta_{ij}\geq 0,\\
    \max\{\Delta_{ij},-x_i\}, & \Delta_{ij}<0,
    \end{cases}
    \label{eq:capped_transfer}
\end{equation}
and the update is
\begin{equation}
    x_i\leftarrow x_i+T_{ij},\qquad
    x_j\leftarrow x_j-T_{ij}.
\end{equation}
Thus, a positive transfer means that agent $j$ pays agent $i$, vice versa, for a negative transfer $i$ pays $j$. The cap enforces nonnegativity and makes $x=0$ absorbing. We call agents with $x_i=0$ inactive.

The exchange rule in Eq.~\eqref{eq:microscopic_transfer} is biased because the resource-dependent score
$x_i^\alpha$
determines the expected direction of transfer.
Only for $\alpha<0$, the transfer favors the poor, driving the system to a uniform wealth distribution, similar to arrested ripening~\cite{Bauermann2025}.
In this paper, we focus on the opposite regime, characterized by an advantage for rich agents.
For $0<\alpha<1$, competitive advantage grows sublinearly with resource; for $\alpha=1$, it grows proportionally; and for $\alpha>1$, it grows superlinearly.
Inequality is therefore generated dynamically: current resources affect future ``bargaining power'', which in turn changes future resource distributions.

\begin{figure}[t]
    \centering
    \includegraphics[width=0.99\linewidth]{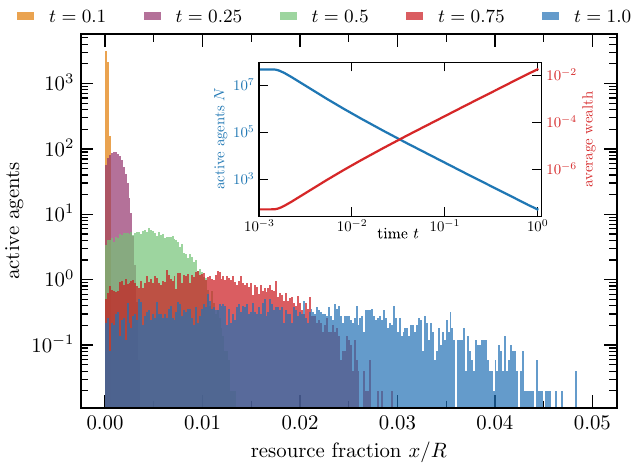}
    \caption{
    Biased asset-exchange dynamics for $\alpha=1/2$, averaged over 50 independent runs.
    Histograms of the resource fraction $x/R$ among active agents at several times show the progressive broadening of the distribution. (Inset) The number of active agents decreases while the average resource of active agents increases.
    }
    \label{fig:agent_model}
\end{figure}

Figure~\ref{fig:agent_model} shows a representative simulation for $\alpha=1/2$. Starting from an almost egalitarian state, small initial fluctuations (of about $1\%$) are amplified.
Many agents go bankrupt and leave the dynamics, while the surviving agents carry an increasing share of the fixed resource $R$. Thus, similar to coarsening, the distribution of active agents that own resources broadens over time and moves towards higher values; compare different time points of the distribution in color. This increase of the mean resource, however, comes with the cost of a decreasing active population (inset).
In contrast to coarsening, however, the distribution broadens faster as time proceeds.

\section{Mean-Field Description}
\label{sec:mean_field}

The biased exchange model has a direct mean-field limit. Let $f(x,t)\,dx$ be the number of active agents with resource between $x$ and $x+dx$, and define the total number of active agents as
\begin{equation}
    N(t)=\int_0^\infty f(x,t)\,dx.
\end{equation}
In a large well-mixed population, the partner of an agent with resource $x$ is sampled from the current empirical distribution. Away from the absorbing boundary, where the cap in Eq.~\eqref{eq:capped_transfer} is inactive for small $\eta$, the conditional mean transfer in one encounter is
\begin{equation}
    \left\langle \Delta x\mid x(t) =x\right\rangle
    =
    \eta
    \left[
        x^\alpha
        -
        \frac{1}{N(t)}\int_0^\infty \xi^\alpha f(\xi,t)\,d\xi
    \right].
    \label{eq:conditional_increment}
\end{equation}

Passing to continuous time and absorbing the encounter rate and $\eta$ into a growth rate $\mu$, assuming each agent trades at a constant rate, the deterministic drift is
\begin{equation}
    \dot{x}=\mu\left[x^\alpha-x_c(t)^\alpha\right],
    \label{eq:growth_law}
\end{equation}
where a critical size $x_c(t)$, defined self-consistently by
\begin{equation}
    x_c(t)^\alpha=
    \frac{\int_0^\infty x^\alpha f(x,t)\,dx}
         {\int_0^\infty f(x,t)\,dx},
    \label{eq:xc_definition}
\end{equation}
emerges.
Agents above $x_c(t)$ grow, while agents below $x_c(t)$ shrink. Thus $x_c(t)$ is the instantaneous unstable fixed point of the size dynamics.

The density is transported by this drift according to continuity equation
\begin{equation}
    \partial_t f(x,t)+\partial_x\!\left(\dot{x}f(x,t)\right)=0,
    \label{eq:continuity_equation}
\end{equation}
with an absorbing boundary at $x=0$. While the total number of active agents is not conserved, because trajectories can reach the absorbing boundary, the total resource is conserved by
\begin{equation}
    R=\int_0^\infty x f(x,t)\,dx.
    \label{eq:global_conservation}
\end{equation}
Indeed, combining Eq.~\eqref{eq:continuity_equation} and Eq.~\eqref{eq:growth_law}, we get
\begin{equation}
    \frac{dR}{dt}=\mu\left[
        \int_0^\infty x^\alpha f(x,t)\,dx-x_c(t)^\alpha\int_0^\infty f(x,t)\,dx\right]=0,
\end{equation}
where the last equality follows from Eq.~\eqref{eq:xc_definition}.

We note that although Eqs.~\eqref{eq:growth_law}--\eqref{eq:continuity_equation} were derived from pairwise exchange, the mean-field model is also relevant in its own right. Not all competitive markets are repeated local contests between pairs. Allocation across stocks within a fixed investment pool, centralized commodity markets, platform-mediated competition, and auction-like mechanisms are organized through market-wide prices, rankings, or clearing rules. In such cases, an agent's growth can depend on its own competitive score relative to a population-level threshold, without any literal microscopic sequence of bilateral exchanges.
The mean-field model therefore captures both the large-population limit of the biased asset-exchange process and a phenomenological description of globally mediated conservative competition.

This model is closely related to LSW-type mean-field theories of ripening~\cite{Lifshitz1961,Wagner1961,Carr1998}. In the linear case $\alpha=1$, Eq.~\eqref{eq:xc_definition} gives $x_c(t)=R/N(t)$, so agents larger than the current mean grow and smaller agents shrink.
In this case, our growth law is equivalent to that of the Carr--Penrose
model~\cite{Carr1998}, $\dot x\propto x/x_c(t)-1$, up to the time change
$dt'=\mu x_c\,dt$, which also describes the continuum limit of the ``very
greedy'' exchange of Ref.~\cite{ispolatov1998}. Here we allow the competitive advantage to scale as a general power of size, which changes the time scales of concentration.

\section{Self-Similar Dynamics}
\label{sec:self_similarity}
We now analyze the long-time scaling behavior of the mean-field equation. Motivated by coarsening theory, we seek solutions of the form
\begin{equation}
    f(x,t)=x_c(t)^{-2}\Phi(y),
    \qquad
    y=\frac{x}{x_c(t)}.
    \label{eq:scaling_ansatz}
\end{equation}
Here, $y$, referred to as a similarity variable, is simply an agent's resource measured in units of the instantaneous critical scale. The scaling ansatz states that, once the resource is measured in similarity variables, the shape of the distribution stops evolving even though $x_c(t)$ itself keeps growing. This ansatz is at the core of LSW theory, where the droplet-size distribution becomes stationary once radii are measured in units of the critical size. $\Phi(y)$ is therefore the time-independent attractor shape, and all of the explicit time dependence is carried by the single scale $x_c(t)$.

The scaling $x_c(t)^{-2}$ is chosen so that $\Phi$ obeys the same conservation of $R$ as in Eq.~\eqref{eq:global_conservation}, i.e.,
\begin{equation}
    \int_0^\infty y \,\Phi(y)\,dy=R.
    \label{eq:scaled_mass}
\end{equation}
The definition of $x_c(t)$ gives the additional self-consistency condition
\begin{equation}
    \int_0^\infty \left(y^\alpha-1\right)\Phi(y)\,dy=0.
    \label{eq:scaled_closure}
\end{equation}
\begin{figure*}[t]
    \centering
    \includegraphics[width=\linewidth]{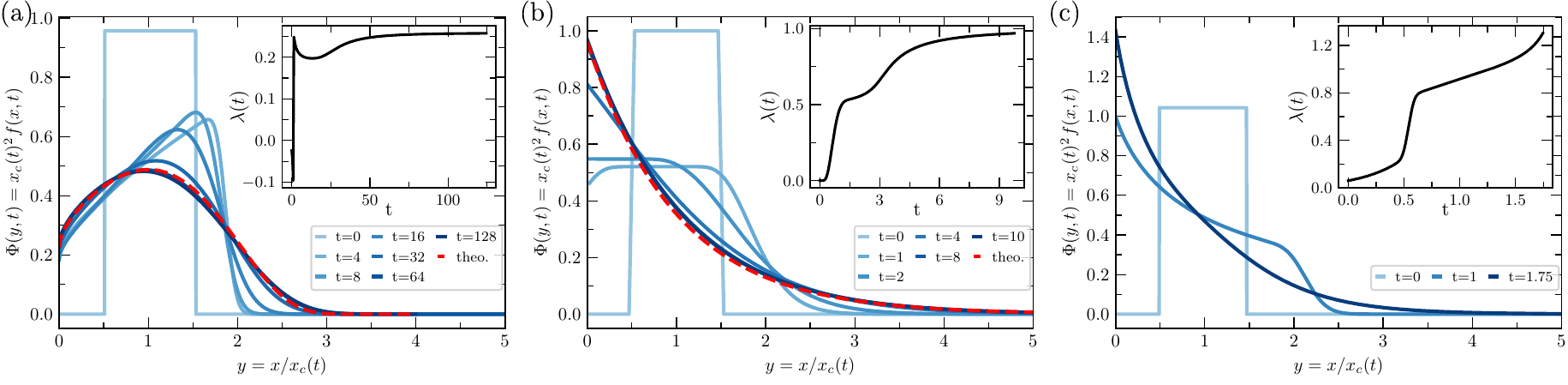}
    \caption{
    Mean-field evolution in similarity variables for three values of the bias exponent.
    The main panels show the rescaled distribution $\Phi(y,t)=x_c(t)^2 f(x,t)$ at different times; insets show the measured $\lambda(t)=\dot{x}_c/(\mu x_c^\alpha)$.
    (a) For $\alpha=1/2$, the solution approaches the selected compact-support profile with $\lamm=1/4$ and $\ym=4$, with $\Phi_{\rm m}(y)=C\left(1-\frac{\sqrt{y}}{2}\right)^{-4}\exp\!\left[-\frac{\sqrt{y}}{1-\sqrt{y}/2}\right]$ where $C=R/(16-32\,e^{2}E_1(2))$ and $E_1(x)$ is the exponential integral.
    (b) For $\alpha=1$, the selected profile is exponential and $\lambda(t)\to 1$.
    (c) For $\alpha=3/2$, the dynamics is sensitive to the far-right tail and approaches a finite-time blow-up regime rather than a stable universal profile.
    Dashed red curves denote the selected theoretical profiles where applicable.
    }
    \label{fig:mean_field_profiles}
\end{figure*}
Substitution of Eq.~\eqref{eq:scaling_ansatz} into Eq.~\eqref{eq:continuity_equation} yields
\begin{equation}
    -\dot{x}_c\left(2\Phi+y\Phi'\right)+\mu x_c^\alpha\partial_y\!\left[(y^\alpha-1)\Phi\right]=0.
    \label{eq:scaled_pde}
\end{equation}
A time-independent shape requires
\begin{equation}
    \lambda \defeq \frac{\dot{x}_c}{\mu x_c^\alpha}
    \label{eq:lambda_definition}
\end{equation}
to be constant. We refer to $\lambda$ as the \textit{scaling rate}. This poses a differential equation for $\dot{x}_c$, whose solution, depending on the value of $\alpha$, reads
\begin{align}
    x_c(t)
    &=
    \left[
        x_c(0)^{1-\alpha}
        +(1-\alpha)\lambda\mu t
    \right]^{1/(1-\alpha)},
    && 0<\alpha<1,
    \label{eq:xc_sublinear}
    \\
    x_c(t)
    &=
    x_c(0)e^{\lambda\mu t},
    && \alpha=1,
    \label{eq:xc_linear}
    \\
    x_c(t)
    &=
    \left[
        x_c(0)^{1-\alpha}
        -(\alpha-1)\lambda\mu t
    \right]^{-1/(\alpha-1)},
    && \alpha>1.
    \label{eq:xc_superlinear}
\end{align}
Importantly, sublinear advantage produces power-law concentration, linear
advantage produces exponential concentration, and superlinear advantage
produces a finite-time blow-up, which for constant $\lambda$ occurs at
\begin{equation}
    t_*=\frac{x_c(0)^{1-\alpha}}{(\alpha-1)\lambda\mu}.
    \label{eq:t_blowup}
\end{equation}
Thus, for $\alpha>1$ a vanishing fraction of agents accumulates all of the conserved resource $R$ within a finite time, leading to a breakdown of the dynamics.

These time scales rely on each agent trading at a constant rate. For a constant
rate per pair, as in the mass-action kinetics of Ref.~\cite{leyvraz2002}, time
is rescaled by a factor $1/x_c(t)$, so that the growth laws above hold with
$\alpha$ replaced by $\alpha-1$, recovering those of Ref.~\cite{leyvraz2002},
while the similarity profiles derived below are unaffected.

We now discuss how to find explicit expressions for the shape function $\Phi(y)$ that obey Eq.~\eqref{eq:scaled_pde}.
Replacing $x_c$ and $\dot{x}_c$ using the constant $\lambda$, as defined in Eq.~\eqref{eq:lambda_definition}, Eq.~\eqref{eq:scaled_pde} admits a whole family of solutions, denoted here as $\Phi_\lambda(y)$.
First, we define the general solutions of this family, that are allowed for the different values of $\alpha$.
Second, in the next section, we discuss how a specific $\lambda$ value gets selected among the allowed ones, leading to an explicit expression of the shape function.

We begin by rewriting Eq.~\eqref{eq:scaled_pde} as
\begin{gather}
    \left[D_\lambda(y)\Phi_\lambda(y)\right]'=\lambda \Phi_\lambda(y),
    \quad \text{with }
    D_\lambda(y) \defeq y^\alpha-1-\lambda y.
    \label{eq:profile_equation}
\end{gather}
With this definition of $D_\lambda(y)$, the growth law Eq.~\eqref{eq:growth_law} can be rewritten in terms of the similarity variable $y$ as
\begin{equation}
    \dot{y}=\mu x_c^{\alpha-1}D_\lambda(y).
    \label{eq:similarity_characteristic}
\end{equation}
Thus, rescaling time by the positive prefactor $\mu x_c^{\alpha-1}$, we can interpret the function $D_\lambda$ as the velocity in similarity variables. This implies that the roots of $D_\lambda$ set a constraint on the support of the shape function $\Phi_\lambda$.
In particular, since $D_\lambda(0)=-1<0$, the first root $\ymax$ is a repelling fixed point of this flow. Thus, no trajectory can reach it in finite (rescaled) time, so that $\ymax$ provides an upper bound for the support of $\Phi_\lambda$.

Whenever the first positive root $\ymax$ exists, the solution on the interval where $D_\lambda<0$ follows by integrating Eq.~\eqref{eq:profile_equation}
\begin{equation}
    \Phi_\lambda(y)=
    -\mathcal{C}_\lambda\frac{\exp\!\left[\lambda\int_0^y\frac{ds}{D_\lambda(s)}\right]
    }{D_\lambda(y)},
    \qquad
    0<y<\ymax.
    \label{eq:formal_profile}
\end{equation}
Here $\mathcal{C}_\lambda>0$ is fixed by the conservation law in Eq.~\eqref{eq:scaled_mass}, and the profile is set to zero beyond its support edge $\ymax$. We discuss three different cases:

\textit{Sublinear bias, $0<\alpha<1$:} $D_\lambda$ is concave, starts at $-1$, and tends to $-\infty$ as $y\to\infty$. It has positive roots when $\lambda$ is smaller than a value $\lamm(\alpha)$ at which the maximum of $D_\lambda$ is zero. This marginal scaling rate $\lamm(\alpha)$ is defined via $D_{\lamm}(\ym)=D'_{\lamm}(\ym)=0$ as
\begin{equation}
    \lamm(\alpha)=\alpha(1-\alpha)^{(1-\alpha)/\alpha},
    \qquad \ym=(1-\alpha)^{-1/\alpha}.
    \label{eq:lambda_c}
\end{equation}
For $0<\lambda<\lamm$, there are two positive roots. The first, $\ymax$, bounds the support, and Eq.~\eqref{eq:formal_profile} gives the profile on $0<y<\ymax$.
At $\lambda=\lamm$, the roots merge and the bound $\ymax = \ym$.
If $\lambda>\lamm$, there is no positive root and the formal solution instead has a tail extending to infinity. Equation~\eqref{eq:profile_equation} can be rewritten as
\begin{equation}
    \frac{d\ln\Phi_\lambda}{dy}
    =-\frac{N_\lambda(y)}{D_\lambda(y)}, \qquad
    N_\lambda(y)=\alpha y^{\alpha-1}-2\lambda.
    \label{eq:log_derivative}
\end{equation}
If $\lambda>\lamm$, this gives $\Phi_\lambda(y)\sim C y^{-2}$, so $\int_0^\infty y\Phi_\lambda(y)\,dy$ diverges logarithmically. There is therefore a formal solution of the ODE, but no physical similarity profile satisfying the finite-mass constraint in Eq.~\eqref{eq:scaled_mass}.

\textit{Linear bias, $\alpha=1$:}
If $\lambda \neq 1$,
Eq.~\eqref{eq:formal_profile} reduces to
\begin{equation}
    \Phi_\lambda(y)=C\,[1-(1-\lambda)y]^{-(2\lambda-1)/(\lambda-1)}.
    \label{eq:al1_Phi}
\end{equation}
Here, for $0<\lambda<1$, $D_\lambda(y)$ has a positive root, thus $\Phi_\lambda(y)$ has finite support with cutoff $\ymax=1/(1-\lambda)$ and Eq.~\eqref{eq:scaled_mass} defines $C$.
For $\lambda>1$ this cutoff does not exist and $\Phi_\lambda(y)$ has infinite support. Nevertheless, the first moment exists, because $\Phi_\lambda(y)$ decays fast enough with exponent $(2\lambda -1)/(\lambda-1)>2$. For $\lambda=1$, there is also no finite root and the solution has infinite support. Here, the solution reads $\Phi_1(y)=Ce^{-y}$, which has a finite first moment.

\textit{Superlinear bias, $\alpha>1$:} The function $D_\lambda$ first decreases and then increases monotonically to $+\infty$, so it has exactly one positive root for every $\lambda>0$. Thus, Eq.~\eqref{eq:formal_profile} gives a profile $\Phi_\lambda(y)$ on $0<y<\ymax$ where the upper bound for the support is implicitly defined via $(\ymax)^\alpha = 1+\lambda \ymax$.

\section{Selection of the Similarity Profile}
\label{sec:selection}

As in LSW theory, Eq.~\eqref{eq:profile_equation} admits a family of similarity profiles, and the selected long-time profile can depend on how the initial distribution approaches its support edge~\cite{Carr1998,Niethammer1999,Niethammer2008}.
Regularizing effects such as diffusion, fluctuations, or discrete encounters can favor a smooth profile~\cite{Giron1998,Meerson1999,Rubinstein2000,Conlon2022}.
Here, we give a selection argument for when a profile is selected in the long time limit. This argument relies on the stability of the dynamics for different $\lambda$ values with respect to perturbation, e.g., numerical noise.
This selection mechanism differs for the three cases, introduced above.

\begin{table*}
\begin{ruledtabular}
\begin{tabular}{cccccc}
 bias & $x_c(t)$, $t>\tsc$ & $\lamm$ & $\ym$ & $\Phim(y)$ & $\Gini_\mathrm{act}$ \\ \hline
 sublinear, $0<\alpha<1$ &
 $\left[ \xsc^{1-\alpha} +(1-\alpha)\lamm\mu (t-\tsc)\right]^{1/(1-\alpha)}$ &
 $\alpha(1-\alpha)^{(1-\alpha)/\alpha}$ &
 $(1-\alpha)^{-1/\alpha}$ &
 $\propto e^{-A_\alpha(z)}/h_\alpha(z)$ &
 Eq.~\eqref{eq:gini-alpha}
 \\
linear, $\alpha=1$ &
 $\xsc\, e^{\mu (t-\tsc)}$ &
 $1$ &
 $+\infty$ &
 $\propto e^{-y}$ &
 $1/2$
 \\
 superlinear, $\alpha>1$ &
diverges at $t_*$ in Eq.~\eqref{eq:t_blowup}, for constant $\lambda$ &
 -- &
 -- &
 not selected &
 --
\end{tabular}
\end{ruledtabular}
\caption{\label{tab:resume} Summary of the self-similar dynamics. For sublinear bias, in the expression of the shape function $\Phim$ we made use of $z=y/\ym\in(0,1)$, $h_\alpha(z)=1-\alpha+\alpha z-z^\alpha$, and $A_\alpha(z)=\alpha\int_0^z dv/h_\alpha(v)$, as in Eq.~\eqref{eq:gini-alpha}; near the edge, $\Phim$ behaves as in Eq.~\eqref{eq:sublinear_phic}. For superlinear bias, no profile is selected, see Sec.~\ref{sec:selection}.}
\end{table*}

\textit{Sublinear bias}: For $0<\alpha<1$, profiles with $\lambda<\lamm$ terminate at the first root $\ymax<\ym$. Increasing $\lambda$ moves this cutoff $\ymax$ to higher values. Thus, any perturbation that places mass just beyond the present cutoff $\ymax$ flows to higher $y$-values because $D_\lambda>0$ there. This drift stops at the marginal profile, where the two roots merge and the cutoff reaches $\ym$, as defined in Eq.~\eqref{eq:lambda_c}.
Thus, the marginal case is the robust limiting profile, and the selected value is the marginal scaling rate $\lamm(\alpha)$.
Expanding Eq.~\eqref{eq:log_derivative} around the cutoff $\ym$ gives
\begin{gather}
    \Phim(y)
    \sim
    C_\mathrm{m}
    (\ym-y)^{-\nu}
    \exp\!\left[
        -\frac{2\ym}{(1-\alpha)(\ym-y)}
    \right],
    \label{eq:sublinear_phic}
    \\
    \nu=
    \frac{10-8\alpha}{3(1-\alpha)}
    \label{eq:sublinear_nu}.
\end{gather}
Hence, the selected profile $\Phim$ vanishes faster than any power of $(\ym-y)$. This selection holds for initial data that vanish faster than any power at the
edge of their support, power-law edges select sub-marginal profiles, as proven
for this class of LSW-like equations~\cite{Niethammer1999,conlon2014}. Note that for $\alpha=1/3$, the shape function matches the classical LSW droplet-size distribution in the volume variable.

\textit{Linear bias}: For $\alpha=1$ and $\lambda <1$, the finite-support cutoff $\ymax=1/(1-\lambda)$ diverges as $\lambda\to \lamm =1$ and the selected profile becomes $\Phim(y)=Ce^{-y}$, as discussed before. In principle, we expect also solutions that are initialized with higher $\lambda$ values to be stable. Therefore, in the linear case, we only have a weak selection mechanism, i.e., acting only for initial distributions whose tails decay sufficiently fast.

\textit{Superlinear bias}: For $\alpha>1$, there exists no marginal scaling rate $\lamm$ and a strong tail sensitivity prevents an analogous universal selection and the profiles describe only a pre-blow-up regime for sufficiently regular initial data.

Figure~\ref{fig:mean_field_profiles} confirms the selection mechanism proposed here. For sublinear ($0<\alpha<1$) and linear ($\alpha=1$) bias, the rescaled distributions approach the selected profiles and $\lambda(t)$ approaches $\lamm$ (panels (a) and (b)).
For superlinear bias, neither the rescaled shape nor $\lambda(t)$ converges (panel (c)).
We summarize the selected scaling laws and profiles for $\alpha\leq1$ in Table~\ref{tab:resume}; for $\alpha>1$, the table reports the finite-time scale divergence without assigning a universal profile.

\section{Dynamics of the Gini coefficient}
\label{sec:gini}
Up to now, we focused on the dynamics of the distribution $f(x,t)$ quantifying the resource belonging to active agents, and its self-similar properties. We now proceed to quantify the inequality associated to such resource distribution, by means of standard economics indicators, i.e., the Lorenz curve $L(u)$ and the associated Gini coefficient $\mathcal{G}$.
In App.~\ref{app:gini}, we give a detailed derivation, here we outline the main results.
In a nutshell, the Lorenz curve plots the cumulative share of the population on the horizontal axis against the cumulative share of resources on the vertical axis. In the case of perfect equality, the Lorenz curve is a straight line bisecting the first quadrant.
The Gini coefficient is twice the area between the bisector of the first quadrant and the Lorenz curve, such that $\mathcal{G} = 0$ represents perfect equality, while $\mathcal{G} = 1$ represents maximal inequality, i.e., one person owns all the resource. In our model inequality increases over time, such that the Lorenz curve and the associated Gini coefficient become dynamical variables, that we indicate as $L(u,t)$ and $\mathcal{G}(t)$, respectively. 

In yard-sale models with a conserved total number of agents without wealth redistribution, it was shown that the Gini coefficient is a Lyapunov functional of the dynamics~\cite{boghosian2015h}.
In our model, the number of agents is not conserved, so that the temporal evolution of the Lorenz curve and the associated Gini coefficient has two contributions.
The first one comes from the change of $f(x,t)$, i.e., the distribution of the resource within the $N(t)$ active agents, while the second stems from the increasing fraction of inactive population, i.e., the population $N_0$ that owns none of the resource $R$. The latter is denoted as $p_0(t)=1-N(t)/N_0$. Since these inactive agents hold zero resource, the Lorenz curve can be defined as
\begin{equation}
L(u,t)=
\begin{cases}
0, & 0\leq u\leq p_0(t),\\[2mm]
L_{\mathrm{act}}\!\left(
  \dfrac{u-p_0(t)}{1-p_0(t)},t
\right), & p_0(t)<u\leq 1,
\end{cases}
\end{equation}
where $u$ is a population fraction and $L_{\mathrm{act}}$ is the Lorenz curve among active agents only. The latter curve is parametrized by the pair $\big(F(x,t),M(x,t)\big)$ with $x\in[0,\infty)$, where
\begin{gather}
F(x,t)=\frac{1}{N(t)}\int_0^x f(x',t)\,dx',\\
M(x,t)=\frac{1}{R}\int_0^x x'f(x',t)\,dx'
\end{gather}
are the fraction of active agents and the fraction of resource held by agents with resource below $x$, i.e., $L_\mathrm{act}(u,t)=M\big(F^{-1}(u,t),t\big)$.
In other words, the presence of inactive agents effectively compresses the active curve into an interval of width $1-p_0$. 
In general, the Gini coefficient is defined via the Lorenz curve as
\begin{gather}
\mathcal{G}(t)
=1-2\int_0^1 L(u,t)\,du  \nonumber
=p_0(t)+(1-p_0(t))\mathcal{G}_{\mathrm{act}}(t)
\\
\text{where } \;
\mathcal{G}_\mathrm{act}(t)
=1-2\int_0^1 L_\mathrm{act}(u,t)\,du \,,
\label{eq:gini-decomposition}
\end{gather}
for a more detailed derivation, see App.~\ref{app:gini}. Here, we focus on the dynamics of $\mathcal{G}(t)$ at late times, i.e., when $f(x,t)$ obeys the self-similar dynamics in Eq.~\eqref{eq:scaling_ansatz}. In this regime, the Gini coefficient of the active population $\mathcal{G}_\mathrm{act}$ becomes time-independent and can be calculated directly from the shape function via
\begin{gather}
\mathcal{G}_{\mathrm{act}}
=
1-\frac{2}{N_\Phi R}
\int_0^\infty \Phi(y)
\left[\int_0^y y'\Phi(y')\,dy'\right]dy,
\\
N_\Phi=\int_0^\infty\Phi(y)\,dy.
\label{eq:gini-active-profile}
\end{gather}
For the marginal profiles with sublinear bias, $0<\alpha<1$, the expression above can be simplified to:
\begin{equation}
\mathcal{G}_{\mathrm{act}}
=
1-\frac{\displaystyle\int_0^1 e^{-2A_\alpha(z)}\,dz}
        {\displaystyle\int_0^1 e^{-A_\alpha(z)}\,dz}\,,
\label{eq:gini-alpha}
\end{equation}
where we have introduced $A_\alpha(z)=\alpha\int_0^z (1-\alpha+\alpha v-v^\alpha)^{-1} dv$ and $z=y/\ym$.
In general, these integrals require numerical evaluation. However, for the limiting case $\alpha =1$, the selected profile is exponential and $\mathcal{G}_{\mathrm{act}}=1/2$.

Thus, within the self-similar regime, the only dynamics of $\mathcal{G}(t)$ stems from the increase in the fraction of inactive agents, with no wealth.
Therefore, we can write the temporal scaling of $\mathcal{G}(t)$ explicitly by denoting $\tsc$ as the start of the self-similar regime, $\Nsc=N(\tsc)$ as the number of active agents, and $\xsc=x_c(\tsc)$ as the critical size at this time.
Combining Eq.~\eqref{eq:gini-decomposition} with the critical-size growth law for $0<\alpha<1$ in Eq.~\eqref{eq:xc_sublinear}, we find
\begin{equation}
\mathcal{G}(t)
=
1-\frac{(1-\mathcal{G}_{\mathrm{act}})\Nsc}{N_0}
\left[1+\frac{(1-\alpha)\lamm\mu(t-\tsc)}{\xsc^{1-\alpha}}
\right]^{1/(\alpha-1)}.
\label{eq:gini-time-sublinear}
\end{equation}
For the linear bias $\alpha=1$, this becomes
\begin{equation}
\mathcal{G}(t)
=
1-\frac{\Nsc}{2N_0}e^{-\mu(t-\tsc)}.
\label{eq:gini-time-linear}
\end{equation}
Thus, within the selected scaling regime, the Gini coefficient approaches 1 with either a power law, for $0<\alpha<1$, or exponentially, for $\alpha =1$.

\section{Comparison with Agent-Based Simulations}
\label{sec:numerics}

\begin{figure}[t]
    \centering
    \includegraphics[width=0.9\linewidth]{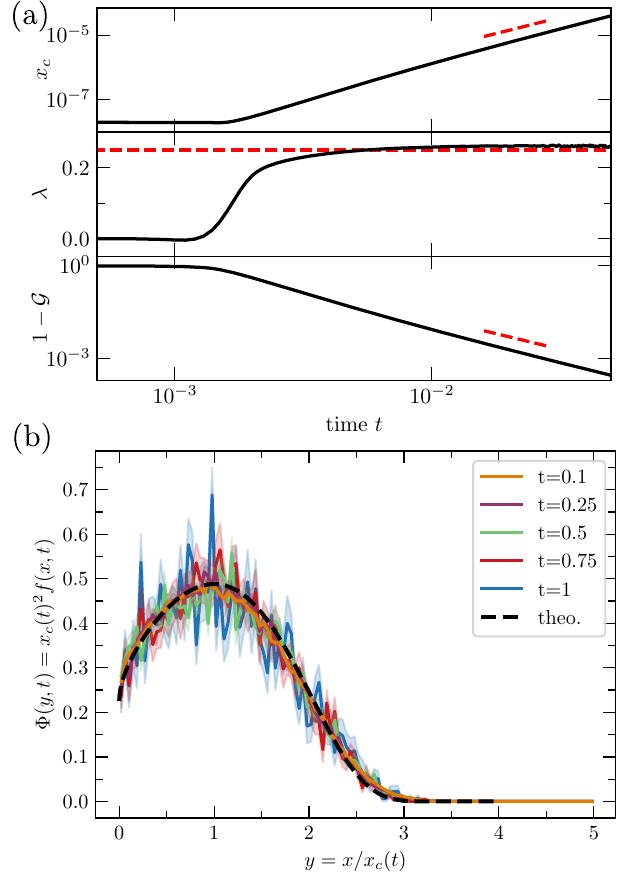}
    \caption{
    Scaling comparison between the biased asset-exchange model and the selected mean-field profile for $\alpha=1/2$.
    (a) The estimated $\lambda(t)$ approaches the selected value $\lamm=1/4$ while the critical size $x_c(t)$ grows, and the Gini coefficient $\Gini$ saturates at 1.
    (b) Rescaled resource distributions collapse onto the selected theoretical profile during the intermediate scaling regime.
    Shaded curves indicate fluctuations across runs.
    }
    \label{fig:agent_scaling}
\end{figure}

We finally return to the biased asset-exchange model and compare it with the mean-field prediction. The comparison for $\alpha=1/2$ is shown in Fig.~\ref{fig:agent_scaling}. We compute the mean-field critical size from the active agents via
\begin{equation}
    x_c(t)=
    \left(
        \frac{1}{N(t)}
        \sum_{i:x_i(t)>0}x_i(t)^\alpha
    \right)^{1/\alpha},
\end{equation}
and estimate
\begin{equation}
    \lambda(t)=\frac{\dot{x}_c(t)}{\mu x_c(t)^\alpha}.
\end{equation}
Furthermore, the Gini coefficient is computed over all $N_0$ agents, inactive ones included, as
\begin{align}
    \Gini(t)&=\frac{1}{2N_0R}\sum_{i,j=1}^{N_0}\left|x_i(t)-x_j(t)\right| \nonumber\\
    &=\frac{2}{N_0R}\sum_{k=1}^{N_0}k\,x_{(k)}(t)-\frac{N_0+1}{N_0},
    \label{eq:gini-abm}
\end{align}
where $x_{(1)}\leq x_{(2)}\leq\dots\leq x_{(N_0)}$ are the resources sorted in increasing order.
The temporal evolution of these three quantities is shown in Fig.~\ref{fig:agent_scaling}~(a).
After an initial transient, the critical size increases as a power-law with exponent $2$ (first panel), while the scaling rate $\lambda(t)$ approaches the selected, marginal mean-field value $\lamm=1/4$ (second panel).
Because the population was initialized close to perfect equality and the biased dynamics leads to a maximal inequality, the Gini coefficient transitions from 0 to 1 (third panel). At late times, $1-\mathcal{G}(t)$ decays as the predicted power law $t^{-2}$, see Eq.~\eqref{eq:gini-time-sublinear}.
In Fig.~\ref{fig:agent_scaling} (b) we show that, at late times, the rescaled resource histograms collapse onto the selected mean-field profile. However, due to the decreasing number of agents, the finite-size fluctuations become more dominant over time.

\section{Conclusion}
\label{sec:conclusion}
We proposed a minimal model combining unfair competition with conserved total resource and agent extinction. These three key ingredients lead to a ripening dynamics in which agents with resource exceeding a critical value get richer while the others get poorer, ultimately leading to a fraction of the population without any resource.
Crucially, the critical resource amount increases in time, reinforcing inequalities among agents. Depending on the strength of the bias, i.e., sublinear, linear or superlinear, the model predicts two qualitatively different dynamical regimes.
On the one hand, continuous resource accumulation within the hands of a few in a self-similar manner; on the other, a winner-takes-all scenario where, 
already at the mean-field level, the dynamics breaks down in finite time.
In the first scenario, we quantified the inequality dynamics analytically, deriving how the critical size and the Gini coefficient scale in time.
Moreover, we derive a family of shape functions at late times, leading to the collapse of the rescaled resource distribution, and propose a mechanism that selects a unique member of this family.
In the winner-takes-all scenario, however, the mean field approximation is less robust, and no unique shape function is selected within the regime prior to the collapse.

The concept of a critical size threshold is widely used in the context of social sciences under the name ``critical mass'' ~\cite{ball2004critical,marwellCriticalMassCollective1993}
where, as in classical nucleation theory~\cite{kelton2010nucleation}, the
threshold is time-independent.
To our knowledge, however, a time-dependent critical size separating growing
from shrinking agents, in the spirit of LSW theory, has not been introduced in
the social sciences. Even scaling theories of migration-driven
growth~\cite{leyvraz2002} and mean-field models of biased
exchange~\cite{ispolatov1998} characterize the dynamics through a typical size,
without relating it to the threshold separating growth from shrinkage.
Our theory explicitly addresses this gap.
To illustrate this change of paradigm in an economic setting, consider what is popularly known as a K-shaped economy, i.e.,
an economy in which a branching point exists, above which rich agents get richer and below which poor agents get poorer. Our framework transforms this static concept into a dynamical one, where the branching point of the K is not a fixed threshold but the critical resource $x_c(t)$ itself, which grows in time. As a consequence, the bar an agent must clear to remain on the ascending branch keeps rising. Agents who once sat comfortably above this threshold can start declining later, not because their own resource fell, in fact they kept on increasing, but because the threshold moved past them.

Similarly, the physics community has largely characterized the scaling in size of complex systems, like cities, firms, etc., at a given time.
Few works, however, highlight instead the scaling in time of the resource or population curves themselves~\cite{complexis16,Ben-Naim2003,leyvraz2002}. Our study shifts the focus from a scale-free stationary size distribution to a scale-free time dynamics of such distributions~\footnote{Except in the special case $\alpha=1$, where $\lambda$ and $ \mu$ combine into a time scale, see Eq.~\eqref{eq:xc_linear}.}. A natural next step is to ask how to model strategies that could counter the advantage of rich agents,  to prevent oligarchy or the hollowing out of small cities.

Our ambitions, however, run into the usual limits of physics applied to society. Echoing Ball's view~\cite{ball2004critical}, a model such as ours can at most show that, if certain structural conditions hold, e.g., extinction coupled to a bias favoring the rich, the collective outcome is a specific, characterizable form of runaway inequality, which can even lead to collapse of the whole dynamics within finite time for a strong enough bias. It cannot tell us, however, where the bias originates, or how to mitigate it in practice.
To put it into the words of Piketty~\cite{piketty2020capital}, ``inequality is neither economic nor technological; it is ideological and political.''

\section{Acknowledgments}
\label{sec:ack}

G.B. acknowledges support from grant JDC2023-051554-I, funded by MCIU/AEI/10.13039/501100011033 and by ESF+. The authors thank Janet van der Graaf Mas for inspiring discussions on the dynamics of cities, René Karadakic for bringing up the US farmland example, Luis Seoane for introducing us to the work of Bruce Boghosian, and Emanuele Cozzo for feedback on the manuscript.

\bibliography{lib}

@book{paretoCoursDeconomiePolitique1896,
  author    = {Pareto, Vilfredo},
  title     = {Cours d'\'economie politique},
  volume    = {1},
  publisher = {F. Rouge},
  address   = {Lausanne},
  year      = {1896}
}

@article{yakovenkoColloquiumStatisticalMechanics2009,
  author  = {Yakovenko, Victor M. and Rosser, J. Barkley},
  title   = {Colloquium: {S}tatistical mechanics of money, wealth, and income},
  journal = {Rev. Mod. Phys.},
  volume  = {81},
  pages   = {1703},
  year    = {2009},
  doi     = {10.1103/RevModPhys.81.1703}
}

@article{Steffen2015,
  author  = {Steffen, Will and Broadgate, Wendy and Deutsch, Lisa and Gaffney, Owen and Ludwig, Cornelia},
  title   = {The trajectory of the {A}nthropocene: {T}he {G}reat {A}cceleration},
  journal = {Anthr. Rev.},
  volume  = {2},
  pages   = {81},
  year    = {2015},
  doi     = {10.1177/2053019614564785}
}

@book{gibratInegalitesEconomiquesApplications1931,
  author    = {Gibrat, Robert},
  title     = {Les in\'egalit\'es \'economiques},
  publisher = {Recueil Sirey},
  address   = {Paris},
  year      = {1931}
}

@article{suttonGibratsLegacy1997,
  author  = {Sutton, John},
  title   = {{G}ibrat's legacy},
  journal = {J. Econ. Lit.},
  volume  = {35},
  pages   = {40},
  year    = {1997}
}

@article{Bouchaud2000,
  author  = {Bouchaud, Jean-Philippe and M\'ezard, Marc},
  title   = {Wealth condensation in a simple model of economy},
  journal = {Physica A},
  volume  = {282},
  pages   = {536},
  year    = {2000},
  doi     = {10.1016/S0378-4371(00)00205-3}
}

@article{gabaixPowerLawsEconomics2009,
  author  = {Gabaix, Xavier},
  title   = {Power laws in economics and finance},
  journal = {Annu. Rev. Econ.},
  volume  = {1},
  pages   = {255},
  year    = {2009},
  doi     = {10.1146/annurev.economics.050708.142940}
}

@article{gabaixZipfsLawCities1999,
  author  = {Gabaix, Xavier},
  title   = {{Z}ipf's law for cities: {A}n explanation},
  journal = {Q. J. Econ.},
  volume  = {114},
  pages   = {739},
  year    = {1999},
  doi     = {10.1162/003355399556133}
}

@article{axtellZipfDistributionUS2001,
  author  = {Axtell, Robert L.},
  title   = {{Z}ipf distribution of {U.S.} firm sizes},
  journal = {Science},
  volume  = {293},
  pages   = {1818},
  year    = {2001},
  doi     = {10.1126/science.1062081}
}

@book{piketty2020capital,
  author    = {Piketty, Thomas},
  title     = {Capital and Ideology},
  publisher = {Harvard University Press},
  address   = {Cambridge, MA},
  year      = {2020}
}

@article{Bouchaud2023,
  author  = {Bouchaud, Jean-Philippe},
  title   = {From statistical physics to social sciences: {T}he pitfalls of multi-disciplinarity},
  journal = {J. Phys. Complex.},
  volume  = {4},
  pages   = {041001},
  year    = {2023},
  doi     = {10.1088/2632-072X/ad104a}
}

@article{CHAKRABORTI2002,
  author  = {Chakraborti, Anirban},
  title   = {Distributions of money in model markets of economy},
  journal = {Int. J. Mod. Phys. C},
  volume  = {13},
  pages   = {1315},
  year    = {2002},
  doi     = {10.1142/S0129183102003905}
}

@article{Boghosian2014,
  author  = {Boghosian, Bruce M.},
  title   = {Kinetics of wealth and the {P}areto law},
  journal = {Phys. Rev. E},
  volume  = {89},
  pages   = {042804},
  year    = {2014},
  doi     = {10.1103/PhysRevE.89.042804}
}

@article{boghosian2015h,
  author  = {Boghosian, Bruce M. and Johnson, Merek and Marcq, Jeremy A.},
  title   = {An {H} theorem for {B}oltzmann's equation for the yard-sale model of asset exchange: {T}he {G}ini coefficient as an {H} functional},
  journal = {J. Stat. Phys.},
  volume  = {161},
  pages   = {1339},
  year    = {2015},
  doi     = {10.1007/s10955-015-1316-8}
}

@inproceedings{complexis16,
  author    = {Boghosian, Bruce M. and Devitt-Lee, Adrian and Wang, Hongyan},
  title     = {The growth of oligarchy in a yard-sale model of asset exchange: {A} logistic equation for wealth condensation},
  booktitle = {Proceedings of the 1st International Conference on Complex Information Systems ({COMPLEXIS})},
  publisher = {SciTePress},
  address   = {Set\'ubal},
  pages     = {187},
  year      = {2016},
  doi       = {10.5220/0005956501870193}
}

@article{Boghosian2017,
  author  = {Boghosian, Bruce M. and Devitt-Lee, Adrian and Johnson, Merek and Li, Jie and Marcq, Jeremy A. and Wang, Hongyan},
  title   = {Oligarchy as a phase transition: {T}he effect of wealth-attained advantage in a {F}okker--{P}lanck description of asset exchange},
  journal = {Physica A},
  volume  = {476},
  pages   = {15},
  year    = {2017},
  doi     = {10.1016/j.physa.2017.01.071}
}

@article{ispolatov1998,
  author  = {Ispolatov, S. and Krapivsky, P. L. and Redner, S.},
  title   = {Wealth distributions in asset exchange models},
  journal = {Eur. Phys. J. B},
  volume  = {2},
  pages   = {267},
  year    = {1998},
  doi     = {10.1007/s100510050249}
}

@article{leyvraz2002,
  author  = {Leyvraz, F. and Redner, S.},
  title   = {Scaling theory for migration-driven aggregate growth},
  journal = {Phys. Rev. Lett.},
  volume  = {88},
  pages   = {068301},
  year    = {2002},
  doi     = {10.1103/PhysRevLett.88.068301}
}

@article{Ben-Naim2003,
  author  = {Ben-Naim, E. and Krapivsky, P. L.},
  title   = {Exchange-driven growth},
  journal = {Phys. Rev. E},
  volume  = {68},
  pages   = {031104},
  year    = {2003},
  doi     = {10.1103/PhysRevE.68.031104}
}

@misc{USDAFarms,
  author = {{USDA Economic Research Service}},
  title  = {Number of {U.S.} farms continues slow decline},
  year   = {2023},
  url    = {https://www.ers.usda.gov/data-products/chart-gallery/gallery/chart-detail/?chartId=58268},
  note   = {accessed May 2026}
}

@misc{RegionaldatenbankDeutschland2026,
  author = {{Statistische \"Amter des Bundes und der L\"ander}},
  title  = {Regionaldatenbank {D}eutschland: {N}umber and size of {G}erman municipalities},
  year   = {2026},
  url    = {https://www.regionalstatistik.de/genesis/online/},
  note   = {accessed May 2026}
}

@misc{FREDTOTALSA,
  author = {{U.S. Bureau of Economic Analysis}},
  title  = {Total vehicle sales [{TOTALSA}], retrieved from {FRED}, {F}ederal {R}eserve {B}ank of {S}t. {L}ouis},
  year   = {2026},
  url    = {https://fred.stlouisfed.org/series/TOTALSA},
  note   = {accessed November 2026}
}

@misc{DOEFact695,
  author = {{U.S. Department of Energy, Vehicle Technologies Office}},
  title  = {Fact \#695: {N}ew car dealerships},
  year   = {2011},
  url    = {https://www.energy.gov/eere/vehicles/fact-695-october-3-2011-new-car-dealerships},
  note   = {data from {NADA} Data 2011; accessed November 2026}
}

@article{voorhees1985theory,
  author  = {Voorhees, Peter W.},
  title   = {The theory of {O}stwald ripening},
  journal = {J. Stat. Phys.},
  volume  = {38},
  pages   = {231},
  year    = {1985},
  doi     = {10.1007/BF01017860}
}

@article{Lifshitz1961,
  author  = {Lifshitz, I. M. and Slyozov, V. V.},
  title   = {The kinetics of precipitation from supersaturated solid solutions},
  journal = {J. Phys. Chem. Solids},
  volume  = {19},
  pages   = {35},
  year    = {1961},
  doi     = {10.1016/0022-3697(61)90054-3}
}

@article{Wagner1961,
  author  = {Wagner, Carl},
  title   = {Theorie der {A}lterung von {N}iederschl\"agen durch {U}ml\"osen ({O}stwald-{R}eifung)},
  journal = {Z. Elektrochem.},
  volume  = {65},
  pages   = {581},
  year    = {1961},
  doi     = {10.1002/bbpc.19610650704}
}

@article{Carr1998,
  author  = {Carr, Jack and Penrose, Oliver},
  title   = {Asymptotic behaviour in a simplified {L}ifshitz--{S}lyozov equation},
  journal = {Physica D},
  volume  = {124},
  pages   = {166},
  year    = {1998},
  doi     = {10.1016/S0167-2789(98)00188-2}
}

@article{Niethammer1999,
  author  = {Niethammer, Barbara and Pego, Robert L.},
  title   = {Non-self-similar behavior in the {LSW} theory of {O}stwald ripening},
  journal = {J. Stat. Phys.},
  volume  = {95},
  pages   = {867},
  year    = {1999},
  doi     = {10.1023/A:1004546215920}
}

@incollection{Niethammer2008,
  author    = {Niethammer, Barbara},
  title     = {On the evolution of large clusters in the {LSW} theory},
  booktitle = {Analysis and Stochastics of Growth Processes and Interface Models},
  publisher = {Oxford University Press},
  address   = {Oxford},
  pages     = {79},
  year      = {2008},
  doi       = {10.1093/acprof:oso/9780199239252.003.0010}
}

@article{conlon2014,
  author  = {Conlon, Joseph G. and Niethammer, Barbara},
  title   = {On global stability for {L}ifschitz--{S}lyozov--{W}agner like equations},
  journal = {J. Stat. Phys.},
  volume  = {154},
  pages   = {1251},
  year    = {2014},
  doi     = {10.1007/s10955-014-0927-9}
}

@article{Giron1998,
  author  = {Giron, B. and Meerson, B. and Sasorov, P. V.},
  title   = {Weak selection and stability of localized distributions in {O}stwald ripening},
  journal = {Phys. Rev. E},
  volume  = {58},
  pages   = {4213},
  year    = {1998},
  doi     = {10.1103/PhysRevE.58.4213}
}

@article{Meerson1999,
  author  = {Meerson, B.},
  title   = {Fluctuations provide strong selection in {O}stwald ripening},
  journal = {Phys. Rev. E},
  volume  = {60},
  pages   = {3072},
  year    = {1999},
  doi     = {10.1103/PhysRevE.60.3072}
}

@article{Rubinstein2000,
  author  = {Rubinstein, B. Y. and Zaltzman, B.},
  title   = {Diffusional mechanism of strong selection in {O}stwald ripening},
  journal = {Phys. Rev. E},
  volume  = {61},
  pages   = {709},
  year    = {2000},
  doi     = {10.1103/PhysRevE.61.709}
}

@article{Conlon2022,
  author  = {Conlon, Joseph G. and Dabkowski, Michael},
  title   = {On global asymptotic stability for the diffusive {C}arr--{P}enrose model},
  journal = {J. Nonlinear Sci.},
  volume  = {32},
  pages   = {75},
  year    = {2022},
  doi     = {10.1007/s00332-022-09838-1}
}

@article{Bauermann2025,
  author  = {Bauermann, Jonathan and Bartolucci, Giacomo and Weber, Christoph A. and J\"ulicher, Frank},
  title   = {Theory of reversed ripening in active phase separating systems},
  journal = {Phys. Rev. Lett.},
  volume  = {135},
  pages   = {148201},
  year    = {2025},
  doi     = {10.1103/f5x9-wp3g}
}

@article{schellingDynamicModelsSegregation1971,
  author  = {Schelling, Thomas C.},
  title   = {Dynamic models of segregation},
  journal = {J. Math. Sociol.},
  volume  = {1},
  pages   = {143},
  year    = {1971},
  doi     = {10.1080/0022250X.1971.9989794}
}

@article{vinkovicPhysicalAnalogueSchelling2006,
  author  = {Vinkovi\'c, Dejan and Kirman, Alan},
  title   = {A physical analogue of the {S}chelling model},
  journal = {Proc. Natl. Acad. Sci. USA},
  volume  = {103},
  pages   = {19261},
  year    = {2006},
  doi     = {10.1073/pnas.0609371103}
}

@article{dallastaAlgebraicCoarseningVoter2008,
  author  = {Dall'Asta, Luca and Galla, Tobias},
  title   = {Algebraic coarsening in voter models with intermediate states},
  journal = {J. Phys. A: Math. Theor.},
  volume  = {41},
  pages   = {435003},
  year    = {2008},
  doi     = {10.1088/1751-8113/41/43/435003}
}

@article{caccioliVoterModelsConserved2013,
  author  = {Caccioli, Fabio and Dall'Asta, Luca and Galla, Tobias and Rogers, Tim},
  title   = {Voter models with conserved dynamics},
  journal = {Phys. Rev. E},
  volume  = {87},
  pages   = {052114},
  year    = {2013},
  doi     = {10.1103/PhysRevE.87.052114}
}

@article{MUNJAL2019182,
  author  = {Munjal, Puja and Kumar, Lalit and Kumar, Sandeep and Banati, Hema},
  title   = {Evidence of {O}stwald ripening in opinion driven dynamics of mutually competitive social networks},
  journal = {Physica A},
  volume  = {522},
  pages   = {182},
  year    = {2019},
  doi     = {10.1016/j.physa.2019.01.109}
}

@book{ball2004critical,
  author    = {Ball, Philip},
  title     = {Critical Mass: How One Thing Leads to Another},
  publisher = {Macmillan},
  year      = {2004}
}

@book{marwellCriticalMassCollective1993,
  author    = {Marwell, Gerald and Oliver, Pamela},
  title     = {The Critical Mass in Collective Action},
  publisher = {Cambridge University Press},
  address   = {Cambridge},
  year      = {1993}
}

@book{kelton2010nucleation,
  author    = {Kelton, Ken and Greer, Alan Lindsay},
  title     = {Nucleation in Condensed Matter: Applications in Materials and Biology},
  series    = {Pergamon Materials Series},
  volume    = {15},
  publisher = {Elsevier},
  address   = {Oxford},
  year      = {2010}
}

\appendix

\section{Lorenz curve and Gini coefficient from the size distribution}
\label{app:gini}

\subsection{Lorenz curve and Gini coefficient for active agents}

In order to calculate the Lorenz curve and the Gini coefficient associated with a given resource distribution, we start by ordering active agents by increasing resource and let $u\in[0,1]$ denote a certain fraction of the active population. The Lorenz curve $L_\mathrm{act}(u,t)$ is the fraction of the total resource held by
the poorest fraction $u$ of active agents. Two extremal cases are $L_\mathrm{act}(u,t)=u$, i.e., perfect equality, and $L_\mathrm{act}(u,t)\to 0$ for
$u<1$, i.e., a single agent accumulates all the resource (winner takes it all). To construct $L_\mathrm{act}(u,t)$ from $f(x,t)$ we need the rescaled population CDF,
\begin{equation}
F(x,t) = \frac{1}{N(t)}\int_0^x f(x',t)\,dx' ,
\label{eq:cdf}
\end{equation}
that maps a resource value $x$ to the fraction of active population $u$ that owns less of that resource, i.e. $F: x \to u= F(x)$. Its inverse, returning $x$ as a function of $u$, is the \emph{quantile function} $x(u,t):=F^{-1}(u,t)$, defined implicitly by
$F\big(x(u,t),t\big)=u$. With $x(u,t)$ in hand we can define the Lorenz curve for the active population
\begin{equation}
L_\text{act}(u,t) = \frac{1}{R}\int_0^{x(u,t)} x' f(x',t)\,dx' = M\big(x(u,t),t\big),
\label{eq:lorenz}
\end{equation}
where we used $R$, the total resource defined in Eq.~\eqref{eq:global_conservation}, as the normalization. In practice, $L_\text{act}$ can be obtained without inverting $F$, by considering the parametric curve, for $x\in[0,\infty)$, of the pair $\big(F(x,t),M(x,t)\big)$ with $M(x,t):=\frac{1}{R}\int_0^x x'f(x',t)\,dx'$.

The Gini coefficient is twice the area between the
equality line and the Lorenz curve,
\begin{equation}
\Gini_\text{act}(t) = 1-2\int_0^1 L_\text{act}(u,t)\,du .
\label{eq:gini-def}
\end{equation}
Substituting Eq.~\eqref{eq:lorenz} into Eq.~\eqref{eq:gini-def} and changing variables from $u$
to $x$ via $du=f(x,t)/N(t)\,dx$, the Gini coefficient of the active population
is expressed as a double integral over $f(x,t)$:
\begin{equation}
\Gini_{\rm act}(t) \;=\; 1-\frac{2}{N(t)\,R}\int_0^\infty f(x,t)
\left[\int_0^x x' f(x',t)\,dx'\right] dx .
\label{eq:gini-active}
\end{equation}

\subsection{Accounting for inactive agents}
\label{sec:gini-inactive}

Equation~\eqref{eq:gini-active} describes inequality among active agents only. The
full population of $N_0$ agents also includes the $N_0-N(t)$ agents that have already hit the
absorbing boundary and sit at $x=0$. These are, by definition, the poorest agents in the system,
and must be included in the Lorenz curve. Let
\begin{equation}
p_0(t) := 1-\frac{N(t)}{N_0}
\end{equation}
be the inactive fraction. In the full-population ordering, the first $p_0(t)$ fraction of agents
are exactly the inactive ones, each holding zero resource, so the Lorenz curve is flat at zero
there: $L(u,t)=0$ for $u\le p_0(t)$. For $u>p_0(t)$, the remaining agents are precisely the
active population, now rescaled to fill the interval $(p_0(t),1]$: writing $w=(u-p_0)/(1-p_0)\in
[0,1]$ for the rank within the active subpopulation,
\begin{equation}
L(u,t) = L_{\rm act}\!\left(\frac{u-p_0(t)}{1-p_0(t)},t\right), \qquad u>p_0(t),
\end{equation}
with $L_{\rm act}$ built from the active population exactly as in Eq.~\eqref{eq:lorenz}
(normalized by $R$, which is unaffected since inactive agents contribute no resource).

Inserting this piecewise curve into Eq.~\eqref{eq:gini-def} and substituting $w=(u-p_0)/(1-p_0)$
in the second piece gives
\begin{align}
\Gini(t) &= 1-2\int_{p_0}^1 L_{\rm act}\!\left(\frac{u-p_0}{1-p_0},t\right) du \notag\\
&= 1-2(1-p_0)\int_0^1 L_{\rm act}(w,t)\,dw \\
&= 1-(1-p_0)\big[1-\Gini_{\rm act}(t)\big], \notag
\end{align}
which rearranges into the exact decomposition
\begin{equation}
\Gini(t) = p_0(t) + \big[1-p_0(t)\big]\,\Gini_{\rm act}(t)\,.
\label{eq:gini-decomp}
\end{equation}
Thus, the inactive fraction $p_0(t)$ is a lower bound for $\Gini(t)$, while inequality among
active agents contributes with weight $1-p_0(t)$.

\subsection{Gini coefficient in the self-similar regime}
\label{sec:gini-shape}

Inserting the scaling form $f(x,t)=x_c(t)^{-2}\Phi(y)$, $y=x/x_c(t)$,
and changing variables $x=y\,x_c(t)$ in all integrals, including their
upper limits, the number of active agents becomes
\begin{equation}
N(t) \to \frac{N_\Phi}{x_c(t)},\quad N_\Phi:=\int_0^\infty\Phi\,dy ,
\end{equation}
and the CDF, Eq.~\eqref{eq:cdf}, depends on $x$ and $t$ only through $y$:
\begin{equation}
F(x,t) \to\frac{1}{N_\Phi}\int_0^{x/x_c}\Phi(y')\,dy':=F_\Phi(y).
\end{equation}
Hence inverting $u=F_\Phi(y)$ we get $ x(u,t)/x_c = y(u)= F^{-1}_\Phi(u)$.
In Eq.~\eqref{eq:lorenz}
the powers of $x_c$ cancel and the upper limit becomes $y(u)$, so
\begin{equation}
\label{eq:Lact_simil}
L_{\rm act}(u) \to \,\frac{1}{R}\int_0^{y(u)}y'\,\Phi(y')\,dy' ,
\end{equation}
which does not depend on time. Likewise, the Gini coefficient of active agents becomes time independent, as can be seen by plugging Eq.~\eqref{eq:Lact_simil} into Eq.~\eqref{eq:gini-def}, recalling that, since $u=F_\Phi(y)$, $du/dy=\Phi(y)/N_\Phi$
\begin{equation}
\Gini_{\rm act}(t)\to
1-\frac{2}{N_\Phi R}\int_0^\infty\Phi(y)\left[\int_0^y y'\Phi(y')\,dy'\right]dy .
\label{eq:gini-infty}
\end{equation}

\end{document}